\documentclass[prl,english,nofootinbib,twocolumn, showkeys]{revtex4-1}

\usepackage[T1]{fontenc}
\usepackage[utf8]{inputenc}
\usepackage{xcolor}

\usepackage{babel}
\usepackage{pmboxdraw}
\usepackage{amstext}
\usepackage{graphicx}
\usepackage{bm}
\usepackage{amsmath}
\usepackage{makecell}
\usepackage{booktabs}
\usepackage[version=4]{mhchem}
\usepackage{comment}
\usepackage{amssymb}
\usepackage{hyperref}

\usepackage{subcaption} 

\begin{document}

\title{Attention to Order: Transformers Discover Phase Transitions via Learnability}

\author{\c{S}ener \"{O}z\"{o}nder}
\affiliation{Institute for Data Science \& Artificial Intelligence, Bo\u{g}azi\c{c}i University, \.{I}stanbul, T\"{u}rkiye}
\email{sener.ozonder@bogazici.edu.tr}

\begin{abstract}

Phase transitions mark qualitative reorganizations of collective behavior, yet identifying their boundaries remains challenging whenever analytic solutions are absent and conventional simulations fail. Here we introduce learnability as a universal criterion, defined as the ability of a transformer model containing attention mechanism to extract structure from microscopic states. Using self-supervised learning and Monte Carlo generated configurations of the two-dimensional Ising model, we show that ordered phases correspond to enhanced learnability, manifested in both reduced training loss and structured attention patterns, while disordered phases remain resistant to learning. Two unsupervised diagnostics, the sharp jump in training loss and the rise in attention entropy, recover the critical temperature in excellent agreement with the exact value. Our results establish learnability as a data-driven marker of phase transitions and highlight deep parallels between long-range order in condensed matter and the emergence of structure in modern language models.

\end{abstract}


\maketitle

Phase transitions are among the most ubiquitous and conceptually unifying phenomena in science.  Whether in the emergence of ferromagnetism from a paramagnet, the onset of superconductivity, the condensation of a fluid or the percolation of connectivity in disordered media, a phase transition marks a qualitative change in collective behavior driven by a single control parameter such as temperature and pressure.  Beyond condensed matter, closely analogous transitions appear in many other fields: epidemics and information cascades in social networks, abrupt regime shifts in ecosystems, forest fire spread and performance cliffs in computational problems.  In all of these examples the competition between coherent, correlated modes and disorder sets a sharp boundary between regimes in which correlations dominate and regimes in which stochasticity washes them out.

Determining where such transitions occur is straightforward only in a handful of exactly solvable models; more often it requires heavy numerical effort, and in many prominent cases both analytical calculations and conventional simulations break down. Disordered systems such as spin glasses defy equilibration, their rugged landscapes trapping standard algorithms. Frustrated quantum magnets host massively degenerate ground states that obscure conventional order parameters. Fermionic models away from special limits encounter the exponential wall of the sign problem and finite-density gauge theories remain opaque to lattice methods for the same reason. In these settings, machine learning offers a different path \cite{Zdeborova2019RevModPhys, Schwab2019, Jared2022, AnimaAnandkumar2023, YasamanBahri2025}. Unsupervised methods applied to Monte Carlo configurations generated from a microscopic Hamiltonian can expose sharp reorganizations in the data that signal a phase boundary, even without prior knowledge of the relevant order parameter. Beyond serving as a workaround when established tools fail, these approaches provide a new lens on emergent collective behavior, capable of revealing hidden structure in regimes that remain resistant to theory and computation alike.

Early explorations of this idea have largely focused on unsupervised methods such as principal component analysis (PCA) \cite{Nieuwenburg2017, Scalettar2017, Wetzel2017, Lyu2022,wang2017machine, wang2018machine}, autoencoders \cite{Scalettar2017, Wetzel2017, Wang2016, Park2025, Acin2020}, artificial neural networks \cite{Carleo2017, Kashiwa2019, Beach2018}, CNNs \cite{Tanaka2017, Melko2017, Beach2018}, SVMs \cite{Liu2019} and deep Boltzmann machines \cite{Morningstar2018}. In these approaches, Monte Carlo spin configurations generated across a range of temperatures are flattened into vectors, with each spin value serving as a feature corresponding to a lattice site. In PCA, the variance is evaluated over ensembles of sample configurations at a given temperature and the leading principal component identifies the direction of maximal fluctuation in that ensemble. The associated eigenvalue grows sharply in the ordered phase at low temperature, reflecting strong correlations among spin degrees of freedom across independent samples, but no strong eigenvalues appear in the disordered phase at high temperature. Autoencoders, on the other hand, extend this logic by compressing spin configurations into a reduced latent space through nonlinear encoding. The emergent low-dimensional representation is expected to capture the phase transition behavior without explicitly constructing an order parameter.

Despite these advances, both PCA and autoencoders face fundamental limitations. Their effectiveness relies on the hope that PCA eigenvalues or the structure of the autoencoder latent space will run parallel to conventional order parameters, yet there is no guarantee that this alignment will occur \cite{Gong2020}. Thus a broadly applicable and physically grounded criterion for identifying phase transitions from microscopic states remains elusive. PCA in particular is restricted to linear correlations, and fails when nonlinear dependencies between lattice degrees of freedom dominate; the inability to capture topological charge correlations such as vorticity in the XY model is a striking example of this limitation \cite{Scalettar2017}. Autoencoders, while nonlinear in principle, suffer from another problem: the content of their latent space is uncontrolled and it is unclear in advance what structures will be encoded. In both approaches, the extracted representations may diverge from true order parameters or other summary quantities derived from microscopic states that ultimately determine the phase boundaries.

\begin{figure*}[!th]
  \centering
  \subcaptionbox{}[0.58\textwidth]{%
    \includegraphics[width=\linewidth]{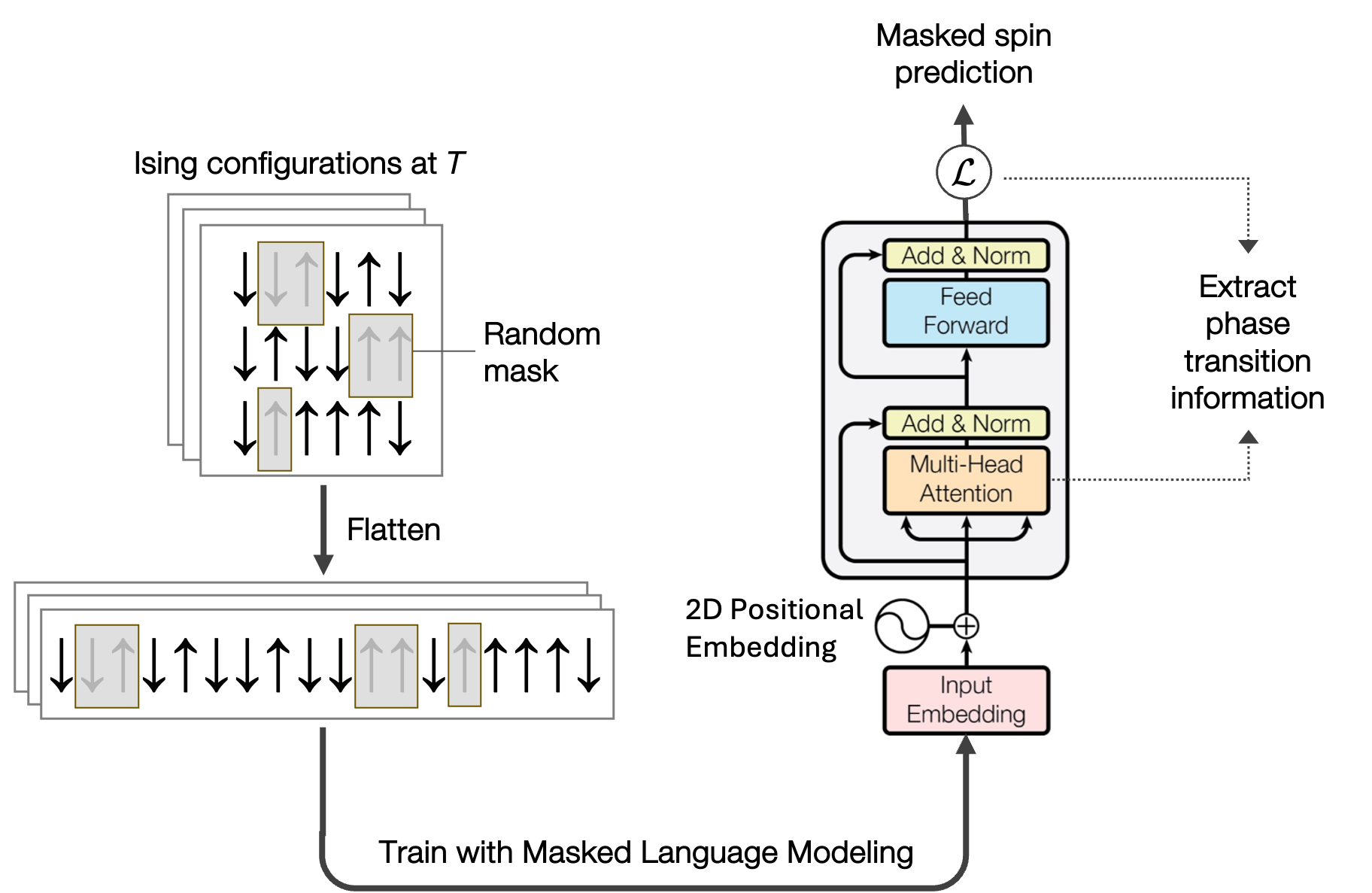}%
  }\hfill
  \subcaptionbox{}[0.42\textwidth]{%
    \includegraphics[width=\linewidth]{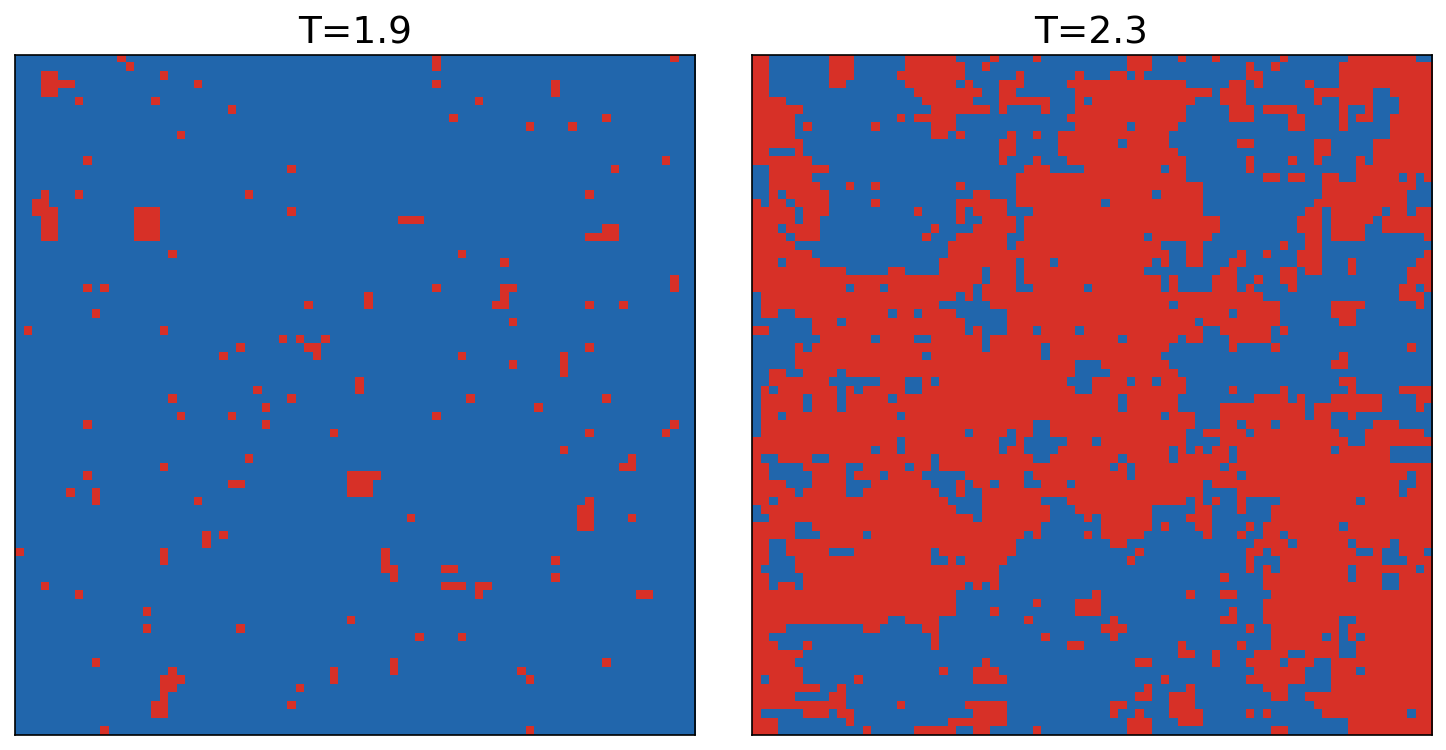}%
  }

  \caption{\textbf{Workflow and sample Ising configurations. a}, Workflow to find $T_c$. \textbf{b}, Ising configurations.}
  \label{fig:workflow}
\end{figure*}

A universal criterion for identifying phase transitions and critical temperatures has yet to be established. In this Letter, we introduce the concept of learnability to fill this gap. The central premise is that data from an ordered phase carry sufficient internal structure to be efficiently captured by a deep learning model, while disordered data resist such compression. Concretely, when a transformer is trained on Monte Carlo configurations from the ordered state, the network rapidly reduces its loss and develops distinct structures in its attention matrices. In contrast, configurations from the high-temperature disordered phase fail to support efficient learning: the training loss remains high after many epochs and the resulting attention patterns lack discernible organization. Thus, the very ability of the model to learn from microscopic states becomes a marker of order–disorder transitions.

Learnability therefore draws a direct link between collective organization and the dynamics of training. Ordered phases, in which long-range correlations establish coherent collective modes, are naturally associated with lower training loss. This mirrors the situation in natural language processing, where large language models succeed because human text exhibits strong correlations: the co-occurrence of words in grammatical sentences provides a learnable structure. When random words are used to form meaningless sequences, the correlations vanish and the model fails to learn. The analogy is immediate: spin degrees of freedom in a lattice behave like tokens in a text corpus and the emergence of order in one domain parallels the emergence of meaning in the other.

Transformers in particular make this connection transparent. Their attention blocks explicitly compute context-dependent correlations between tokens, a mechanism that uncovers long-range dependencies in text. When applied to spin configurations, the same machinery reveals collective correlations among microscopic degrees of freedom. In both cases, meaning (or order) emerges only through the coherent interplay of long-range dependencies across the system. The distinction between ordered and disordered phases is thus imprinted directly into the network’s attention structure, with learnability serving as a bridge between statistical physics and modern deep learning. The method requires only raw Monte Carlo configurations and is agnostic to the existence of a conventional order parameter, making it suitable for frustrated magnets, spin glasses, or other strongly correlated systems where traditional tools fail. 

\begin{figure*}[!th]
  \centering
  \subcaptionbox{}[0.53\textwidth]{%
    \includegraphics[width=\linewidth]{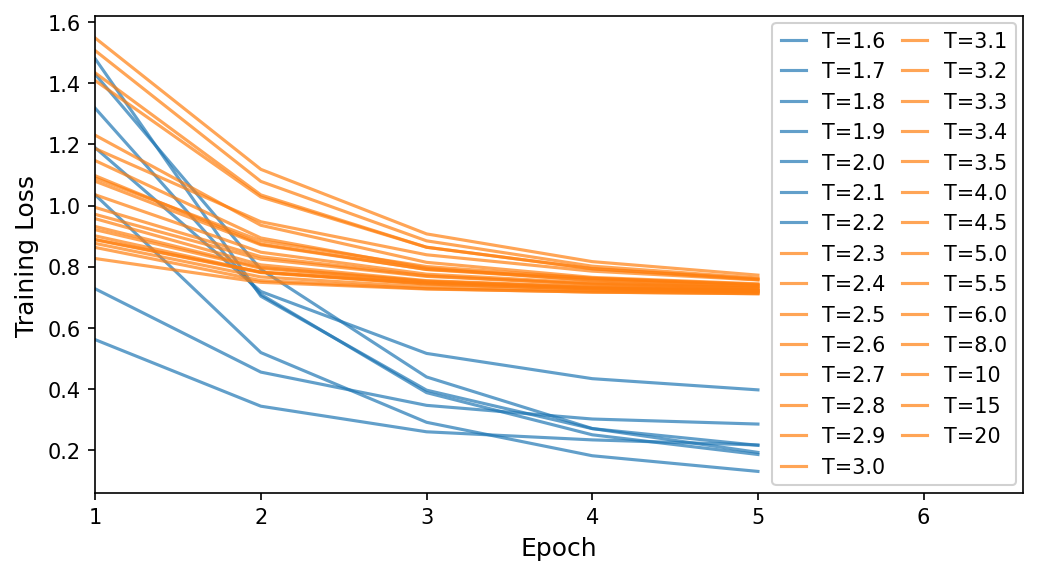}%
  }\hfill
  \subcaptionbox{}[0.42\textwidth]{%
    \includegraphics[width=\linewidth]{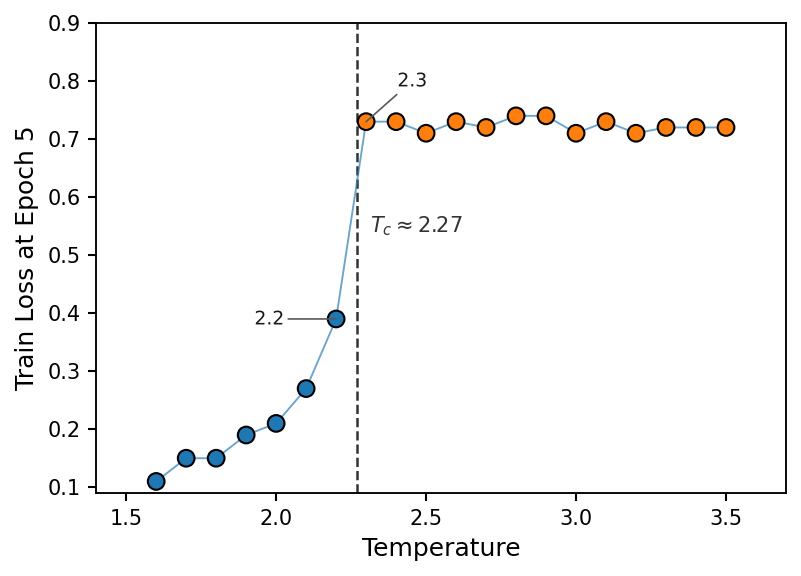}%
  }

  \caption{\textbf{Training losses. a}, Training curves. \textbf{b} Training losses at epoch 5.}
  \label{fig:losses}
\end{figure*}

To illustrate the concept of learnability, we consider the two-dimensional Ising model with nearest-neighbour spin interactions, described by the Hamiltonian
\begin{equation}
H = - J \sum_{\langle i,j \rangle} s_i s_j,
\end{equation}
where each spin takes values $s_i \in \{-1,1\}$. For the ferromagnetic case ($J=1$), the critical temperature is known analytically,
$T_c = 2/\ln(1+\sqrt{2}) \approx 2.27$,
which separates the low-temperature ferromagnetic (ordered) phase from the high-temperature paramagnetic (disordered) phase. Monte Carlo sampling generates spin configurations at a desired temperature according to the Boltzmann distribution,
\begin{equation}
P(\{s_i\}) \propto \exp\!\left(-\frac{H}{T}\right),
\end{equation}
providing the raw data for our study. Specifically, we generate 100 independent $80 \times 80$ configurations at each of 29 temperatures spanning the range $T \in [1.6,20]$. For each bundle of 100 configurations corresponding to a fixed temperature, an encoder-only transformer is trained from scratch. The resulting loss values and attention-head entropies are then compared across temperatures, enabling the unsupervised identification of the critical point at the phase boundary. Crucially, the transformer has no access to temperature labels: phase boundaries are inferred directly from the microscopic states in an unsupervised way.

Transformers in large language models learn to encode each token in text as a high-dimensional embedding vector, initialized randomly and progressively shaped during training to reflect statistical relations among words. These embeddings capture co-occurrence and word order, forming the basis for semantic representation. In the present context, the vocabulary consists of the two spin values $s_i = \{-1,1\}$. Unlike text that requires 1D positional encoding, however, 2D Ising spin configurations necessitate a 2D positional encoding to represent 2D neighborhood relations on the lattice. Training is performed in a self-supervised manner: during each epoch, 6\% of the spins in every configuration are randomly replaced by a special mask token and the objective is to predict the masked values. This masked-language-modeling scheme tunes the attention weights and feed-forward layers so that the network captures the correlations characteristic of a given temperature, effectively learning the language of the Ising model.

The workflow for identifying the critical temperature is illustrated in Fig.~\ref{fig:workflow}a. Representative Ising configurations at $T=1.6$ and $T=20$ are displayed in Fig.~\ref{fig:workflow}b. Each $80\times80$ configuration is flattened into a one-dimensional sequence and fed to the transformer. Binary cross-entropy loss is used to train the network to predict the masked spins. Because no temperature labels are provided, the model is compelled to exploit intrinsic statistical regularities of the configurations. At high temperatures, $T \gg T_c$, thermal fluctuations wash out correlations imposed by the Hamiltonian, and the transformer can only learn short-range structure, yielding relatively high loss. At low temperatures, $T  < T_c$, magnetization emerges as spins become correlated over long distances and the transformer achieves markedly lower loss. This mirrors the situation in natural language learning: when a text corpus consists of random words, a transformer fails to reduce its training loss, whereas for a corpus containing meaningful sentences, the model captures semantic structure by decreasing the loss. In this work, we demonstrate that this notion of learnability, the ability of a deep learning model to exploit collective order, is a reliable criterion for separating ordered and disordered phases.

\begin{figure*}[!th]
  \centering
  \subcaptionbox{}[\textwidth]{%
    \includegraphics[width=\textwidth]{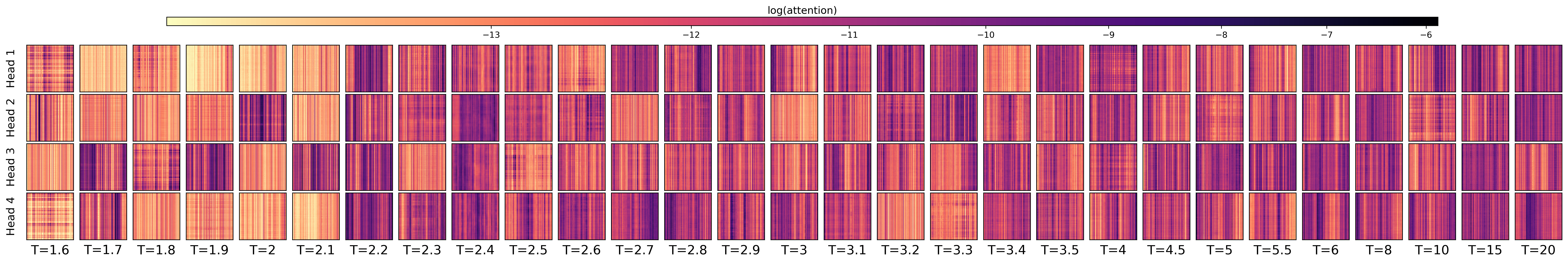}%
  }\\[0.5em]  

  \subcaptionbox{}[0.49\textwidth]{%
    \includegraphics[width=\linewidth]{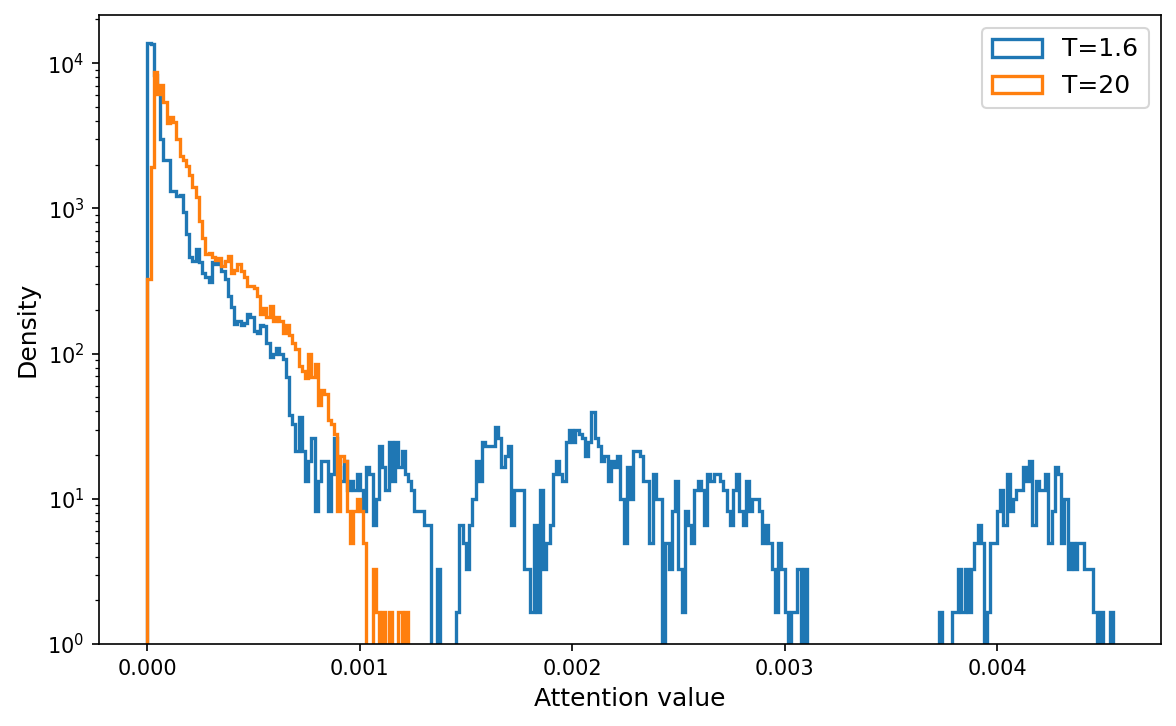}%
  }\hfill
  \subcaptionbox{}[0.49\textwidth]{%
    \includegraphics[width=\linewidth]{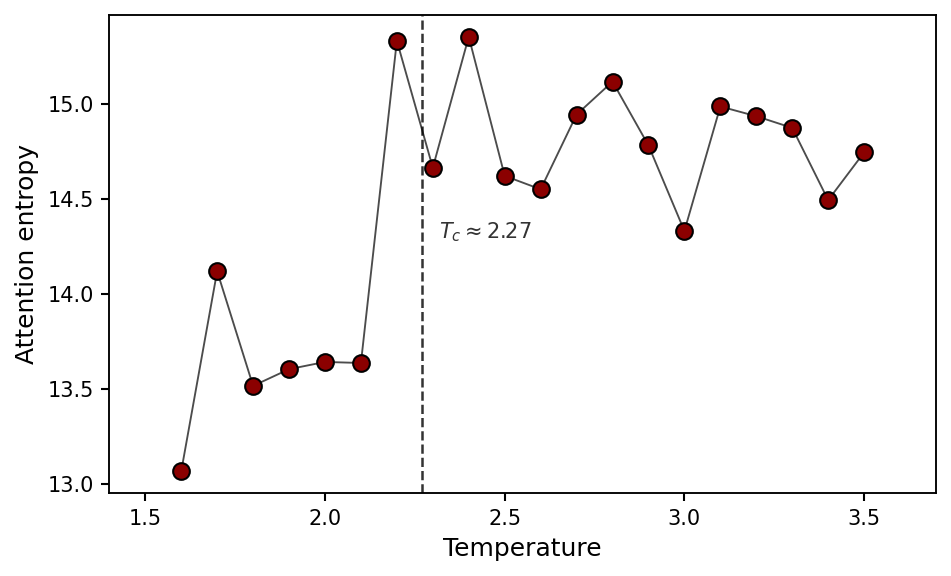}%
  }

  \caption{\textbf{Attention heads after training. a}, Attention heads. \textbf{b}, Attention histograms at two different phases, $T=1.6$ and $T=20$. 
           \textbf{c}, Attention entropy vs. temperature}
  \label{fig:attention}
\end{figure*}

We use an encoder-only transformer trained in a masked-language-modeling (MLM) objective on flattened $80\times 80$ Ising configurations; hence the sequence length is $S=6400$. At each iteration, a fraction $p=0.06$ of sites are replaced by the mask token; targets for unmasked positions are excluded from the loss. Tokens are embedded into $\mathbb{R}^{d}$ with $d=128$ and augmented with a two-dimensional sinusoidal positional encoding, constructed as the tensor product of 1D sinusoidal embeddings along the horizontal and vertical lattice axes. The encoder stack comprises one layer with four heads, followed by a feed-forward network of width 512 with ReLU activation functions. Training uses binary cross-entropy, optimized with Adam at learning rate $2\times 10^{-5}$ for 5 epochs. For each temperature $T$ (29 values in $[1.6,20]$, 100 configurations per $T$), a fresh model is trained from scratch. This configuration suffices to learn long-range dependencies when present while remaining light enough to train repeatedly across temperatures.

After pretraining with the masked-language-modeling objective, the transformer is not further fine-tuned for any downstream task specific to locating $T_c$. Instead, we propose two unsupervised methods for retrieving the critical temperature: (i) identifying the temperature at which the final training loss exhibits a sharp jump, and (ii) examining the attention-head structure by calculating attention entropy and detecting an abrupt change.

In the first method, the final training loss is recorded after training the transformer on 100 independent Ising configurations at a fixed temperature and this procedure is repeated for all sampled temperatures. The temperature dependence of the final loss then reveals a clear separation between phases. As shown in Fig.~\ref{fig:losses}a, for $T>2.2$ the models converge to uniformly high loss values, reflecting low learnability. In contrast, for $T \leq 2.2$ the loss is consistently smaller, indicating that ordered spin configurations contain structured correlations that the transformer can learn. The critical temperature extracted from this separation is in excellent agreement with the exact analytical value $T_c \approx 2.27$. This behavior is illustrated further in Fig.~\ref{fig:losses}b, which plots the loss after 5 epochs for each temperature and reveals a sharp rise between $T=2.2$ and $T=2.3$.

In the second method, we examine the $6400\times6400$ attention matrices obtained at all temperatures, as shown in Fig.~\ref{fig:attention}a. These matrices are evaluated after training by forward-passing four Ising configurations from the same temperature with random masks. Both axes correspond to the flattened spin positions and each entry quantifies how strongly two spins attend to each other, i.e., their learned correlations. Because the attention heads are initialized with independent random weights at the start of training, each head learns a different structure from the configurations. As temperature increases, the color scale of the attention plots changes abruptly between $T=2.1$ and $T=2.2$, marking a transition between distinct learnability regimes associated with the ferromagnetic and paramagnetic phases.

Figure~\ref{fig:attention}b presents the distributions of attention values, averaged over the four heads, for two representative temperatures in the ordered ($T=1.6$) and disordered ($T=20$) phases. At $T=1.6$, the strong spin–spin correlations manifest as a higher density of large attention values, while at $T=20$ such large weights are absent. Because attention matrices are normalized, the concentration of mass in high values at low $T$ necessarily leaves the remainder distributed at very low values, producing a bimodal distribution. In contrast, in the disordered phase the absence of large weights results in the bulk of the distribution accumulating around intermediate values. This difference explains why the attention maps in Fig.~\ref{fig:attention}a appear visually lighter in the ordered phase, as the visualization is dominated by the many small values.

To quantify the spin–correlation information encoded in the attention matrices, we use a Shannon–entropy on the trained attention heads. For each temperature $T$ and each attention head $A^{(h)}$ where $h=1,\dots,4$, we flatten $A^{(h)}$ to $a^{(h)}=\mathrm{vec}\!\left(A^{(h)}\right)$ and define a probability vector
\begin{equation}
p^{(h)}_i \;=\; \frac{a^{(h)}_i}{\sum_{j=1}^{S^2} a^{(h)}_j}\,, \qquad i=1,\dots,S^2,
\end{equation}
where $S=6400$ is the sequence length. The per-head attention entropy is then
\begin{equation}
H^{(h)}(T) \;=\; - \sum_{i=1}^{S^2} p^{(h)}_i \,\log\!\big(p^{(h)}_i \big),
\label{eq:attn-entropy}
\end{equation}
and we aggregate across four heads by a simple average,
\begin{equation}
\overline{H}(T) \;=\; \frac{1}{4}\sum_{h=1}^{4} H^{(h)}(T).
\end{equation}
This quantity measures how dispersed the attention mass is over all site pairs: when correlations are long-ranged and structured (ordered phase), attention concentrates on a sparse set of entries, producing a lower entropy; when correlations are washed out (disordered phase), attention spreads more uniformly and the entropy increases. We find that $\overline{H}(T)$ exhibits a pronounced increase across the phase boundary, as shown in Fig.~\ref{fig:attention}c, providing an unsupervised, information-theoretic marker of the transition. 

We have demonstrated that phase boundaries can be uncovered directly from microscopic configurations using self-supervised learning, without access to temperature labels during training. By introducing learnability as a physically motivated criterion, manifested here through a sharp change in training loss and an abrupt rise in attention entropy, we established a robust correspondence between ordered phases and enhanced learnability: distinct phases exhibit distinct learnability patterns. Applied to the two-dimensional ferromagnetic Ising model, our approach recovers the known critical temperature ($T_c \approx 2.27$) from raw Monte Carlo data via two complementary, unsupervised diagnostics.

The precision of the discovered $T_c$ can be systematically improved by increasing dataset size and refining the temperature grid. Because the method relies only on configurations and not on an explicit order parameter, it naturally extends to other lattices, interaction ranges, and Hamiltonians, including frustrated or disordered systems, offering a general route to chart phase diagrams where conventional tools are challenged. Conversely, insights gained from modeling physical correlations illuminate how information flows through attention and feed-forward blocks, providing a principled avenue toward explainability in large language models and deep learning more broadly \cite{Tegmark2017, Ganguli2020, Galitski2024, Tiwary2024, Zdeborova2025Attn}.

The Monte Carlo generated Ising configurations and the code for training the transformer and Python files for analyses can be found on the \href{https://github.com/senerozonder/Ising-transformer}{GitHub} page.

\section{Acknowledgements.} 

The author is supported by TÜB\.{I}TAK under grant no. 120F354. Computing resources used in this work were provided by Barcelona Supercomputing Center. 



\bibliographystyle{naturemag}
\bibliography{references}

\end{document}